\documentclass[journal]{IEEEtran}
\ifCLASSINFOpdf
\usepackage[pdftex]{graphicx}
\graphicspath{figures/}
\DeclareGraphicsExtensions{.pdf,.jpeg,.png}
\else
\usepackage[dvips]{graphicx}
\graphicspath{{figures/}}
\DeclareGraphicsExtensions{.eps}
\fi
\usepackage[cmex10]{amsmath}
\usepackage{bm}
\usepackage{hyperref}
\usepackage{mathtools}
\usepackage{helvet}
\usepackage{xcolor}
\usepackage{cite}
\usepackage{amsfonts}
\usepackage{subfigure}

\usepackage{url}
\begin{document}
\title{Leveraging Probabilistic Switching in Superparamagnets for Temporal Information Encoding in Neuromorphic Systems}

\author{Kezhou~Yang, Dhuruva Priyan~G M, and~Abhronil~Sengupta,~\IEEEmembership{Member,~IEEE}
\thanks{Manuscript received April, 2022.}
\thanks{The authors are with the School
of Electrical Engineering and Computer Science, Department of Materials Science and Engineering, The Pennsylvania State University, University Park,
PA 16802, USA. E-mail: sengupta@psu.edu.\\
The work was supported in part by the National Science Foundation grants ECCS \#2028213 and CCF \#1955815 and by Oracle Cloud credits and related resources provided by the Oracle for Research program.
}}
\maketitle
\begin{abstract}
Brain-inspired computing - leveraging neuroscientific principles underpinning the unparalleled efficiency of the brain in solving cognitive tasks - is emerging to be a promising pathway to solve several algorithmic and computational challenges faced by deep learning today. Nonetheless, current research in neuromorphic computing is driven by our well-developed notions of running deep learning algorithms on computing platforms that perform deterministic operations. In this article, we argue that taking a different route of performing temporal information encoding in probabilistic neuromorphic systems may help solve some of the current challenges in the field. The article considers superparamagnetic tunnel junctions as a potential pathway to enable a new generation of brain-inspired computing that combines the facets and associated advantages of two complementary insights from computational neuroscience -- how information is encoded and how computing occurs in the brain. Hardware-algorithm co-design analysis demonstrates $97.41\%$ accuracy of a state-compressed 3-layer spintronics enabled stochastic spiking network on the MNIST dataset with high spiking sparsity due to temporal information encoding.
\end{abstract}

\maketitle


\section{Introduction}
Deep learning has undergone unprecedented growth in the past decade and has witnessed success in a plethora of applications. However, with scaling complexity of the problem space and with the ever-growing dimensions of data, computational expenses to train and implement such Artificial Intelligence (AI) systems have also grown beyond limits. Driven by this motivation, ``neuromorphic computing" attempts to decode the operation of the biological brain by mimicking the core functionalities in the underlying algorithms and hardware substrate. In particular, we focus on the more bio-plausible ``spiking" neural/synaptic computing models in this text due to its promise of enabling low-power, asynchronous ``compute only when needed" neuromorphic hardware. We will refer to such a computing model as ``Spiking Neural Networks" (SNNs) for the remainder of this text. While SNNs have shown initial promise as a low-power, event-driven alternative computing paradigm, significant challenges remain from both the algorithms and hardware perspective to ensure scalability in terms of key performance metrics like recognition accuracy, hardware power, energy and area efficiency. Most prior studies have used smaller sub-problems or have converted non-spiking Deep Neural Networks (DNNs) to SNNs \cite{sengupta2019going} - a non-optimal approach in demonstrating the abilities of SNNs. Currently, SNNs remain very similar to non-spiking networks with the analog neural computation in DNNs distributed as binary information over time in the case of a spiking neuron - with the temporal aspect remaining largely unexploited. This has significantly limited SNN efficiency in large-scale problems \cite{davies2021advancing}.

In order to address these limitations, we formulate our solution against two complementary backdrops:

\noindent $\bullet$ \textbf{\textit{Information Encoding (Goal - Enhanced Sparsity and Reduced Latency):}} The vast majority of SNN algorithm formulations have been based on rate coding \cite{severa2019training,shrestha2018slayer} where the neuron output is encoded in the spike rate, i.e. the total number of spikes generated in a sufficiently long time duration. However, in temporal-encoding, the precise time duration required to spike is believed to encode the neuron output information. The principal advantages of using temporal encoding \cite{guo2021neural} for modelling spiking behavior are multiple. Since information is now transmitted in precise spike timings instead of the signal rate, such neural codes can be sparse and much faster to avoid temporal-averaging effect. 

\noindent $\bullet$ \textbf{\textit{Computing Paradigm (Goal - State-Compressed Hardware):}} The computing perspective is motivated by a bottom-up hardware viewpoint that emerging technologies like spintronics exhibit stochastic switching behavior (due to thermal noise) at room temperature, specially at aggressively scaled dimensions \cite{camsari2017stochastic,sengupta2016magnetic}. The potential benefits of such a computing framework from the hardware implementation perspective is that they allow multi-level neural/synaptic state compression to single bit (in turn, leading to scaled device implementations) due to the additional probabilistic encoding of information. However, such stochastic SNNs have been mostly utilized in the rate encoding framework. 

In order to leverage the benefits of increased information capacity in SNNs for enhanced power, latency and energy metrics and simultaneously to utilize the advantages of state-compressed hardware enabled by these nanomagnetic devices, the article explores a device-algorithm co-design approach -- where we explore the implementation of spintronics enabled stochastic SNNs bearing temporal domain encoding of information. Section II discusses basic device preliminaries of magnetic tunnel junction devices. Section III outlines the novel device physics enabling the dynamic temporal control of the stochastic magnetization dynamics that are leveraged in Section IV to formulate algorithms for stochastic SNNs with temporally encoded spikes. Recognition accuracy and spiking sparsity advantages for fully connected network architectures on the MNIST dataset are reported in Section IV. Section V concludes the paper with potential future research directions.

\section{Magnetic Tunnel Junction (MTJ) as a Stochastic Computing Element}
Magnetic Tunnel Junction is a fundamental device building block of spintronic hardware systems. A typical MTJ consists of two ferromagnetic layers and a sandwiched oxide layer. One of the ferromagnetic layers is called “pinned layer” (PL) because its magnetization direction is “pinned” and does not change during operation. The other ferromagnetic layer is called “free layer” (FL) since its magnetization can be switched freely by an external stimuli like spin current or magnetic field. The state of the device is determined by the relative orientation of the two ferromagnetic layers. The device is in “anti-parallel” (AP) / “parallel” (P) state if the two ferromagnetic layers have opposite / same magnetization direction. The device possesses a higher resistance in AP state than in the P state. Energy barrier height determined by device volume and anisotropy stabilizes the two states.

Landau-Lifshitz-Gilbert (LLG) equation with a spin torque term is used to characterize the probabilistic switching of an MTJ device\cite{slonczewski1989conductance},
\begin{equation}
\label{llg}
\frac {d\widehat {\textbf {m}}} {dt} = -\gamma(\widehat {\textbf {m}} \times \textbf {H}_{eff})+ \alpha (\widehat {\textbf {m}} \times \frac {d\widehat {\textbf {m}}} {dt})+\frac{1}{qN_{s}} (\widehat {\textbf {m}} \times \textbf {I}_s \times \widehat {\textbf {m}})
\end{equation}
in which $\widehat {\textbf {m}}$ is the FL magnetization unit vector, $\gamma= \frac {2 \mu _B \mu_0} {\hbar}$ is the gyromagnetic ratio, $\alpha$ is Gilbert\textquoteright s damping ratio, $\textbf{H}_{eff}$ is the effective magnetic field, $N_s=\frac{M_{s}V}{\mu_B}$ is the number of spins in free layer of volume $V$ (where $M_{s}$ is saturation magnetization and $\mu_{B}$ is Bohr magnetron), $q$ is the charge of a single electron and $\textbf{I}_{s}$ is the input spin current. Thermal noise is included by adding an additional thermal field, $\textbf{H}_{thermal}=\sqrt{\frac{\alpha}{1+\alpha^{2}}\frac{2K_{B}T_{K}}{\gamma\mu_{0}M_{s}V\delta_{t}}}G_{0,1}$, where $G_{0,1}$ is Gaussian distribution with zero mean and unit standard deviation, $K_{B}$ is Boltzmann constant, $T_{K}$ is the absolute temperature and $\delta_{t}$ is the simulation time-step.

\begin{figure}[!t]
\centering
\includegraphics[width=2.2in]{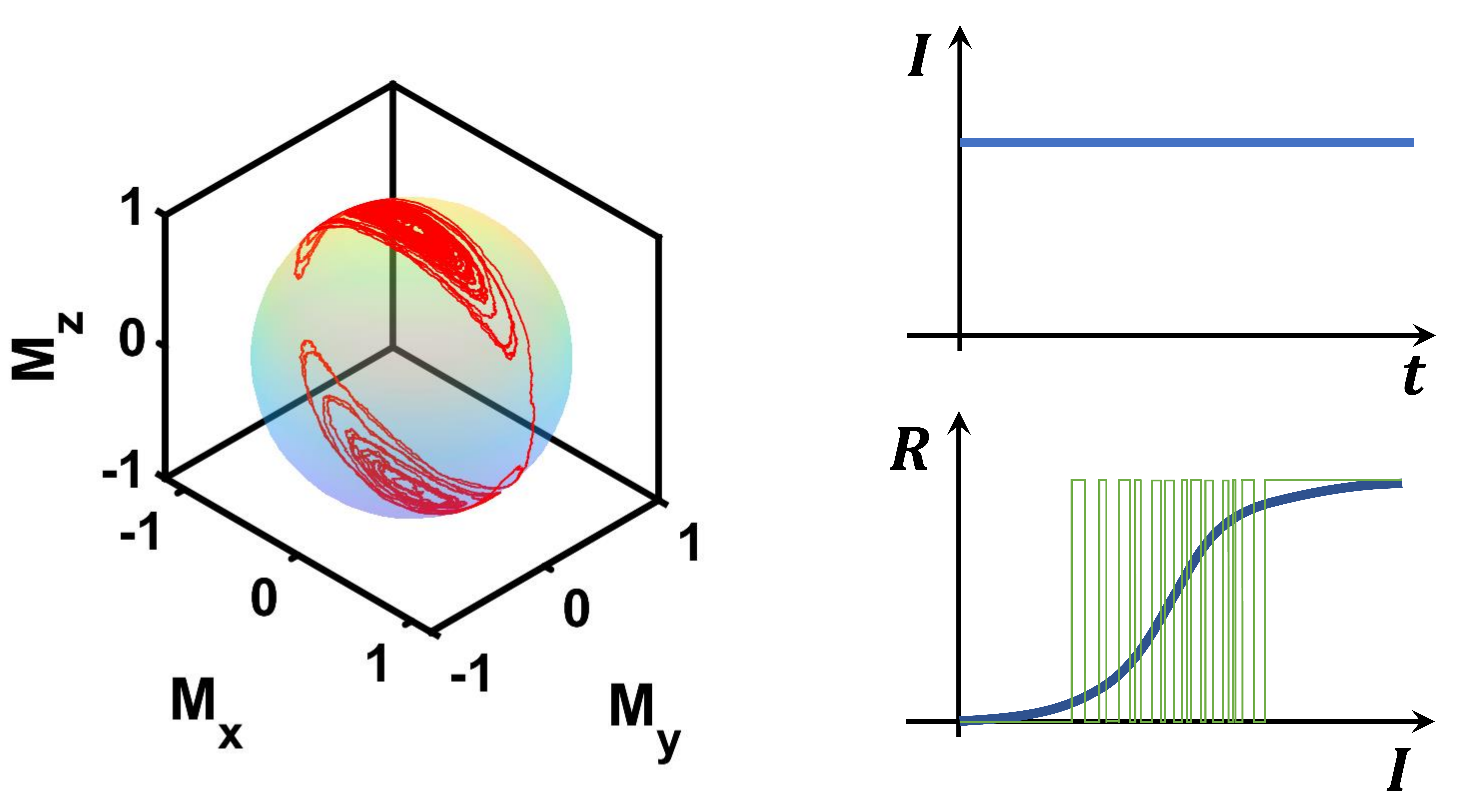}
\caption{\textbf{Device Preliminaries:} Magnetization components for a magnet with anisotropy along the z-direction is shown during a switching process. For a superparamagnetic device, the switching is spontaneous as shown by the noisy switching characteristics. However, the device lifetimes can be modulated by the external current stimuli, $I$, resulting in a sigmoid device switching rate, $R$, variation with external current magnitude. The green transients represent the plot without time averaging.}
\label{fig1}
\vspace{-4mm}
\end{figure}
Recently there has been a lot of interest in superparamagnetic devices for unconventional computing. In essence, these are aggressively scaled nanomagnetic MTJs in the sub-$10K_{B}T_{K}$ barrier height regime where the magnet loses its non-volatility and does not need to be triggered by a pulse for state transitions (see Fig. 1). The thermal noise becomes significant and is large enough to overcome the barrier height, resulting in spontaneous stochastic switching behavior. However, the metastable state transitions can be modulated by an external current and the time-averaged response of the device, $R = \frac{\tau_{AP}}{\tau_{P}+\tau_{AP}}$ ($\tau_{P}$ and $\tau_{AP}$ are device lifetimes in the P and AP states respectively) has a non-linear sigmoid response that can be utilized for stochastic spiking neuron functionalities \cite{sengupta2016probabilistic}. The main advantage of transitioning to a superparamagnetic system would lie in the faster operating speeds and asynchronous operation \cite{liyanagedera2017stochastic}. However, careful peripheral circuitry design, sensitivity to noise and variations remain open challenges \cite{liyanagedera2017stochastic}. In addition to neuromorphic applications \cite{sengupta2018stochastic,sengupta2016magnetic,srinivasan2016magnetic,liyanagedera2017stochastic,roy2018perspective,behin2016building}, stochasticity inherent in magnetic devices (superparamagnets or higher barrier height magnets) have been leveraged to implement true random number generators \cite{vodenicarevic2017low}, and even for other unconventional computing platforms like Ising computing, quantum-inspired algorithms, combinatorial optimization problems, on-chip temperature sensors, among others \cite{camsari2017stochastic,sengupta2017magnetic,shim2018biased,camsari2019p}. 

While the intrinsic temporal dynamics of superparamagnets have been utilized in certain applications like Ising computing, the vast majority of neuromorphic SNN applications have primarily leveraged the superparamagnetic device characteristics in the rate encoding regime, i.e. the continuous-time dynamic behavior of superparamagnets have been ignored and the time-averaged behavior has been used from the computing perspective. This leads us to the question - \textit{Can the unique probabilistic switching behavior of superparamagnetic devices be utilized for temporal information encoding in stochastic SNNs?} 
\vspace{-2mm}

\section{Leveraging the Dynamic Temporal Behavior of MTJs}
In order to design a magnetic device where the intrinsic physics is able to support temporal information encoding, one needs to precisely control the device lifetimes $\tau_P$ and $\tau_{AP}$. This is difficult in a superparamagnet under sole external current stimulation. As shown in Fig. 1, the external current magnitude and direction controls the time averaged firing rate of the device and both the device lifetimes get modulated together with change in the external current magnitude. 

However, as explained in Eq. (1), the magnetization dynamics is a function of both external current and external magnetic field which opens up the possibility of tuning the two device lifetimes by two separate independent control knobs. When an external ``write" voltage is applied to the MTJ (resulting in spin-torque) along with an external magnetic field,  the lower MTJ resistance in the P state results in much larger modulation of $\tau_{P}$ than  $\tau_{AP}$ due to an external voltage. Consequently, the external spin current can be used to control $\tau_{P}$. On the other hand, the magnetic field can be used to tune $\tau_{AP}$ by manipulating the energy profile. In this manner, under certain conditions \cite{zink2019independent}, independent control of $\tau_{P}$ and $\tau_{AP}$ can be realized by adjusting the externally applied magnetic field and current. Recent experiments \cite{zink2018telegraphic} and theoretical modelling \cite{zink2019independent} have shown that such a controlling scheme can be realized in a CoFeB MTJ stack within a range of applied field and current.
\vspace{-4mm}
\begin{figure}[htp]
  \centering
  \subfigure[]{\includegraphics[scale=0.027]{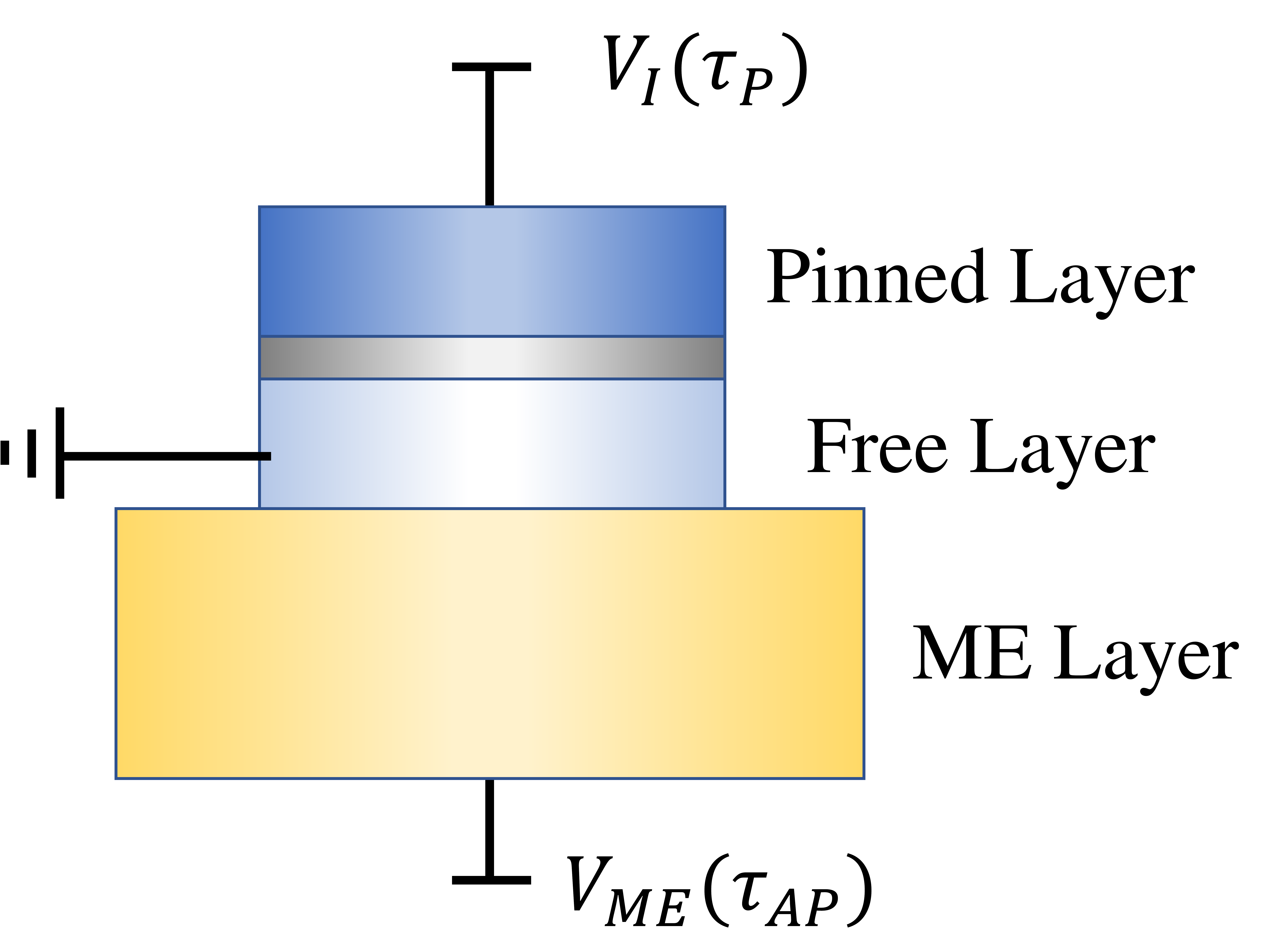}}\quad
  \subfigure[]{\includegraphics[scale=0.032]{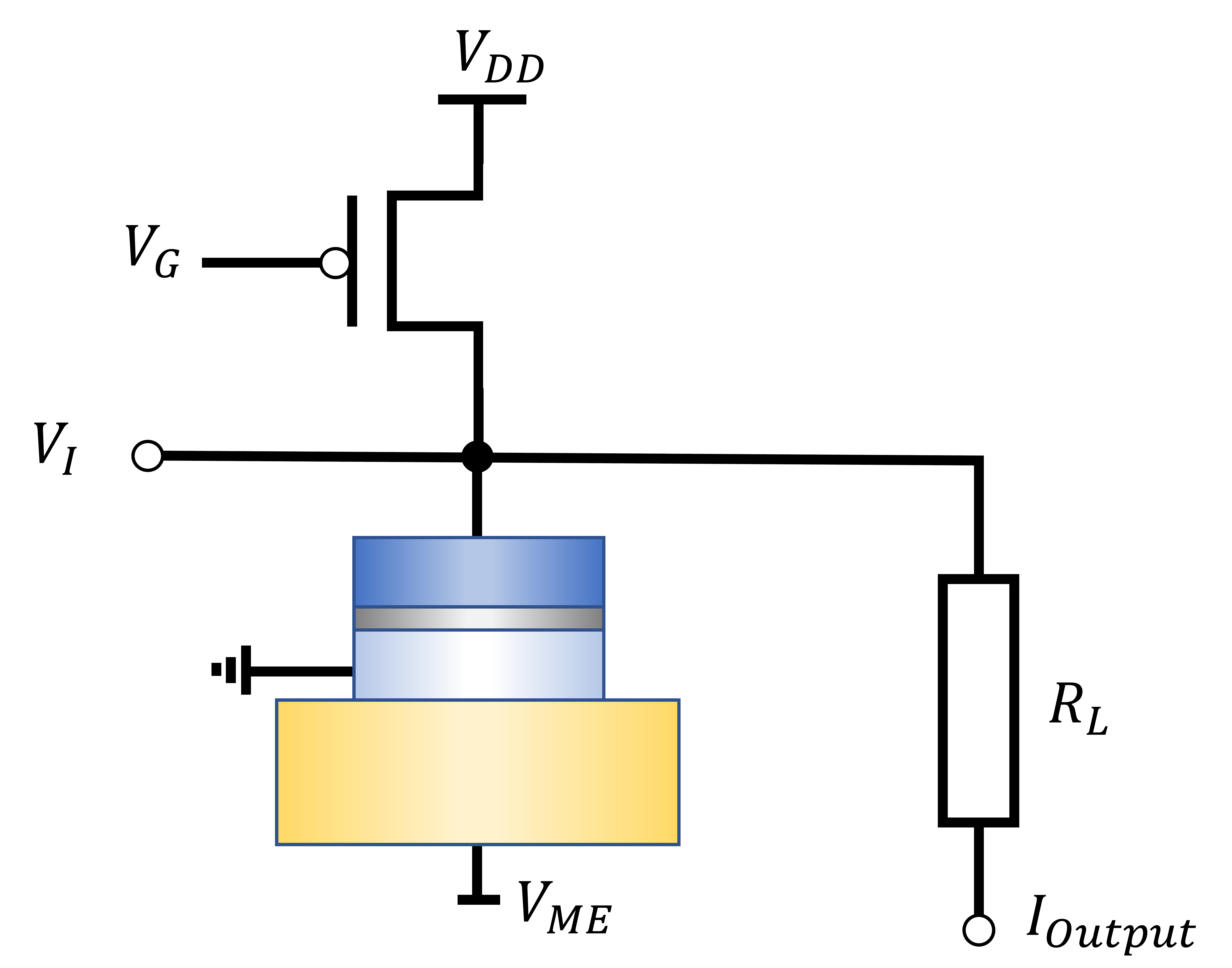}}\quad
  \subfigure[]{\includegraphics[scale=0.05]{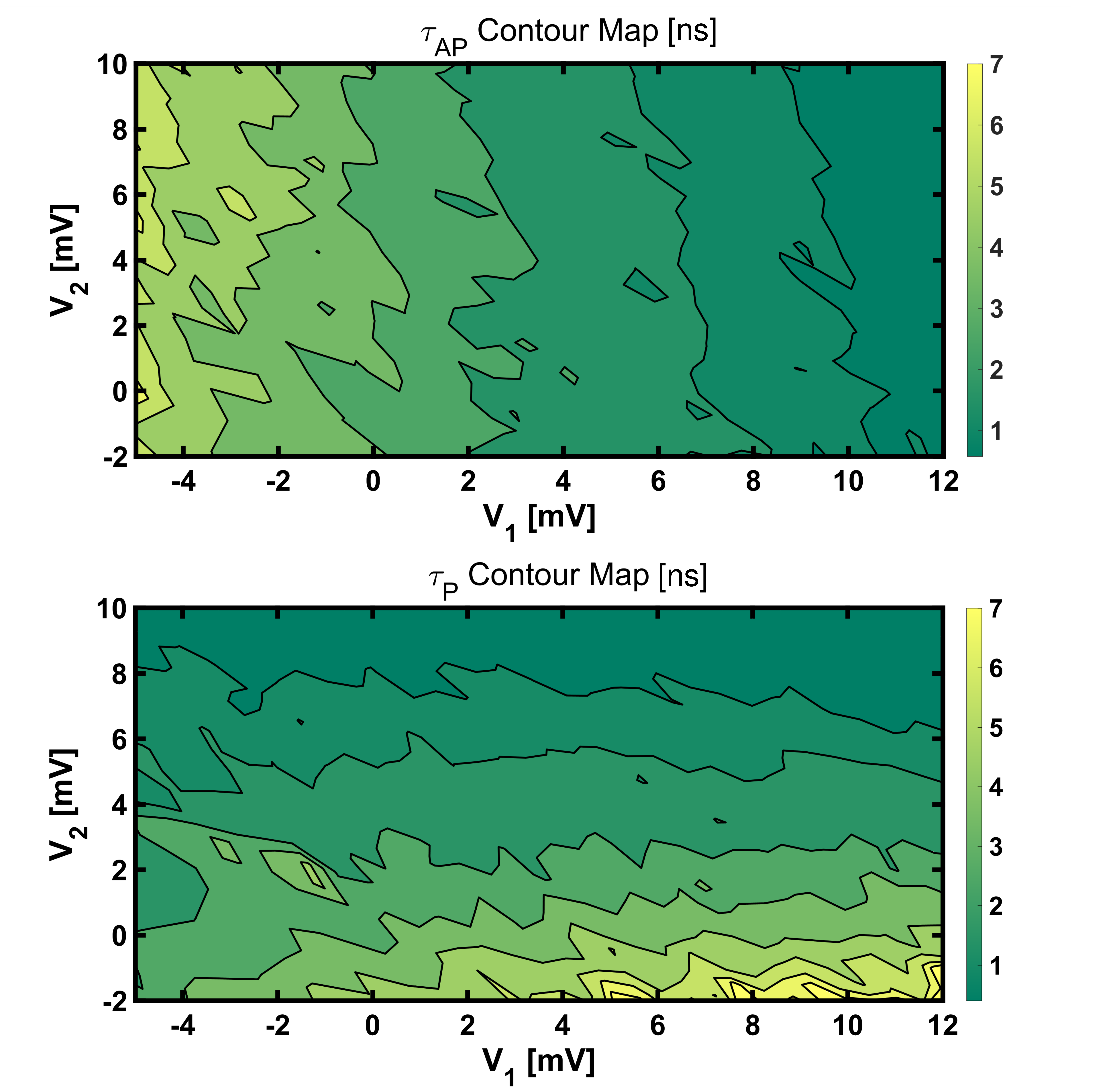}}
  \caption{\textbf{Stochastic computing and temporal information encoding in MTJs:} (a) Concept of magneto-electric MTJ device \cite{yang2020stochastic}, driven by two independent inputs - (1) Voltage, $V_{ME}$, applied across the ME-oxide modulates lifetime $\tau_{AP}$, (2) Voltage, $V_{I}$, applied across the MTJ modulates $\tau_{P}$. (b) Circuit design to detect spikes. $I_{Output}$ indicates the MTJ state. (c) Contour map of $\tau_{AP}$ and $\tau_{P}$ versus external voltage inputs $V_{1}$ and $V_{2}$  \cite{yang2020stochastic}. The horizontal and vertical nature of the contour lines indicate independent control of the device lifetimes.}
\end{figure}

However, on-chip external magnetic field control of nanoscale devices is not promising from the effect of scalability and power consumption \cite{yang2020stochastic}. A potential alternative path can be to design novel device structures exploiting emerging devices physics like the magnetoelectric effect \cite{cheng2018recent}. Recent work \cite{yang2020stochastic} explored a three-terminal magnetoelectric (ME) MTJ device concept where voltage applied across a ME layer ($V_{ME}$) lying underneath the MTJ was used to mimic the effect of an effective magnetic field while voltage across the MTJ stack ($V_{I}$) was used to induce an external spin current, as is shown in Fig. 2(a). ME effect was modelled by considering the effect of an external magnetic field acting on the magnet, whose magnitude is directly proportional to the applied voltage \cite{nikonov2014benchmarking,chakraborty2018design}, with the proportionality factor ($\alpha_{ME}$) being a material property. The device modelled at room temperature ($300K$) has an elliptic ferromagnetic layer, the size of which is $17 nm$ in width, $42.5 nm$ in length and $0.8 nm$ in thickness. Tunnel magnetoresistance (TMR) ratio of the device is $200 \%$. The saturation magnetization is $750 KA/m$. Gilbert damping ratio is chosen to be $0.0122$. The ME layer has a thickness of $5 nm$ and ME constant of $5 \times 10^{-9} s/m$ \cite{liyanagedera2017stochastic,chakraborty2018design}. The device state can be detected by a circuit shown in Fig. 2(b). The transistor working in saturation region provides a constant current, $I_{Total}$. $V_I$ is the input voltage applied to the MTJ. The MTJ resistance modulates the current flowing through the MTJ, $I_{MTJ}$, leading to the control of current flowing through the load resistance $R_L$. As a result, the output current, $I_{Output}=I_{Total}-I_{MTJ}$, will be an indicator of the MTJ state. While some amount of inter-dependency of the device lifetimes is observed, it can be shown through device characterizations that the device lifetime modulation can be made truly independent by a simple transformation of the external voltages to a different bases $<V_1,V_2>$ which can be mapped to the device inputs $<V_{ME},V_{I}>$ through the relation \cite{yang2020stochastic},
\begin{equation}
\begin{pmatrix}
V_1 \\
V_2
\end{pmatrix} = \begin{pmatrix}
\cos \alpha & \cos \beta \\
\sin \alpha & \sin \beta 
\end{pmatrix}^{-1} \begin{pmatrix}
V_{ME} \\
V_I
\end{pmatrix}
\label{conversion}
\end{equation}
where, $\alpha, \beta$ represent the slopes of the contour lines for $\tau_{P}$ and $\tau_{AP}$ variation against $<V_{ME},V_{I}>$. For more details, interested readers are referred to Ref. \cite{yang2020stochastic}. The transformed input $V_{1}$ ($V_{2}$) only controls $\tau_{AP}$ ($\tau_{P}$) independently, as shown by the horizontal/vertical contour lines in Fig. 2(b). It is worth mentioning here that this transformation can be achieved in hardware by simple voltage summer circuits since $\alpha$ and $\beta$ are constants.

\begin{figure*}[htp]
  \centering
  \subfigure[]{\includegraphics[scale=0.038]{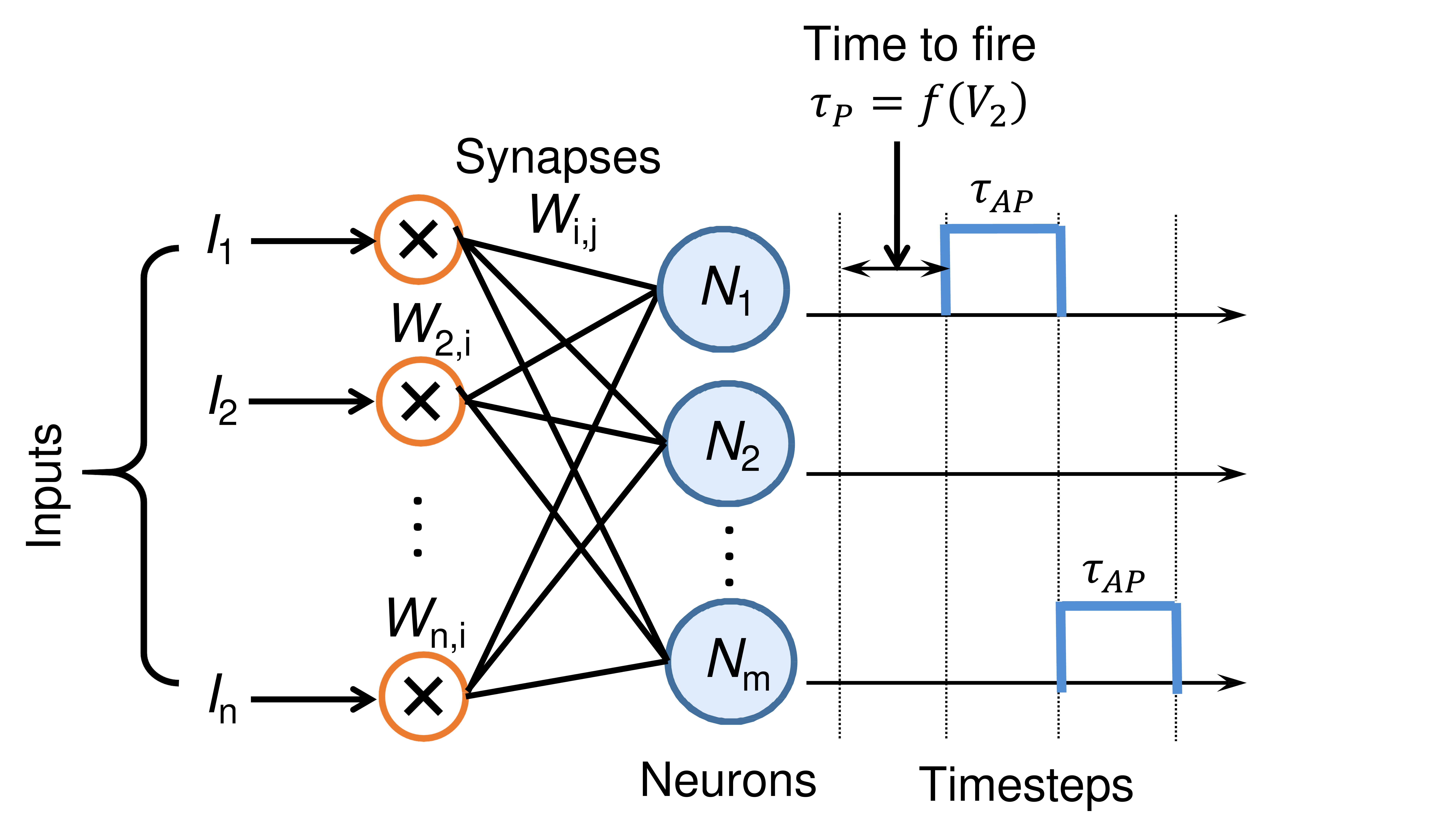}}\quad
  \subfigure[]{\includegraphics[scale=0.036]{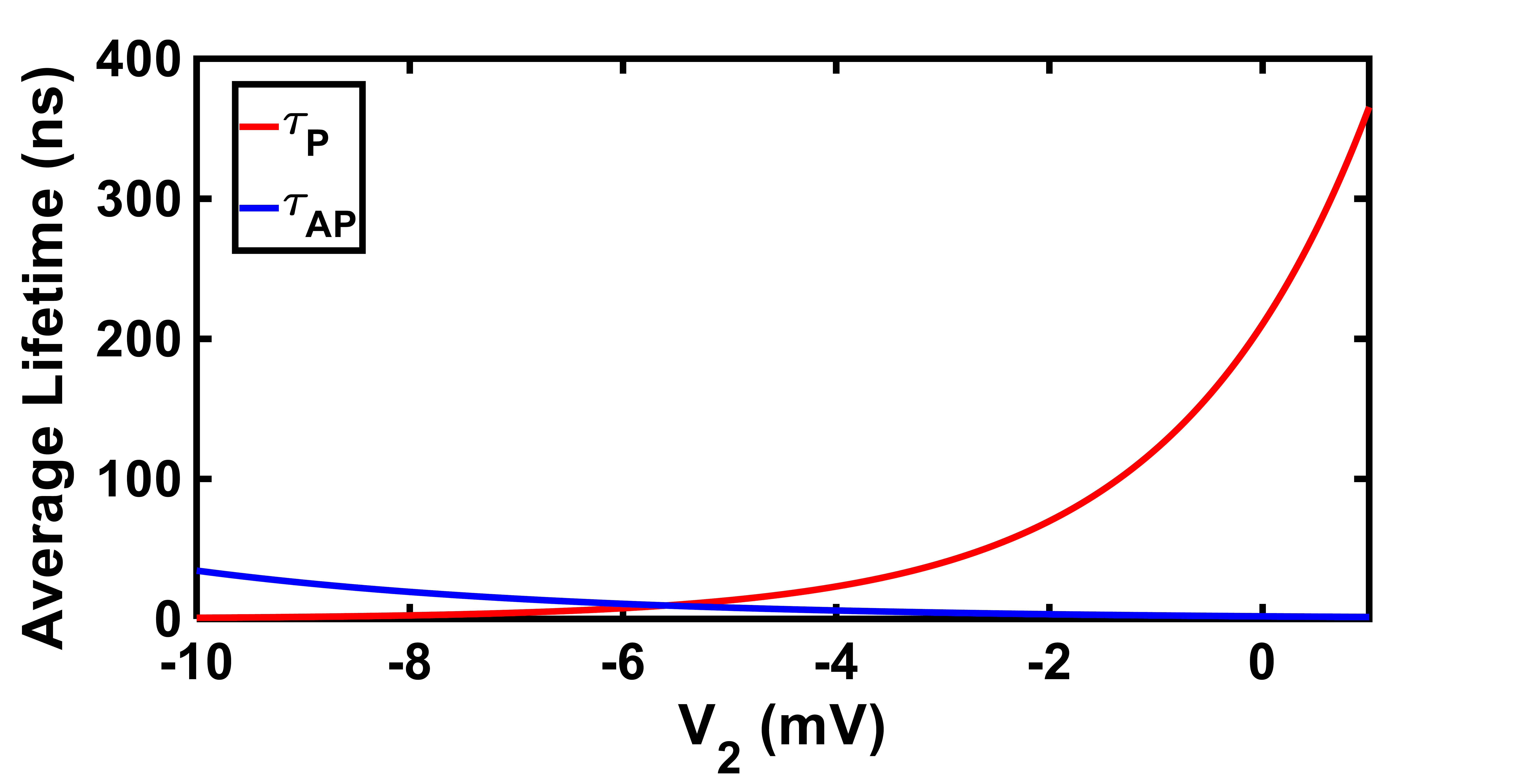}}\quad
  \subfigure[]{\includegraphics[scale=0.034]{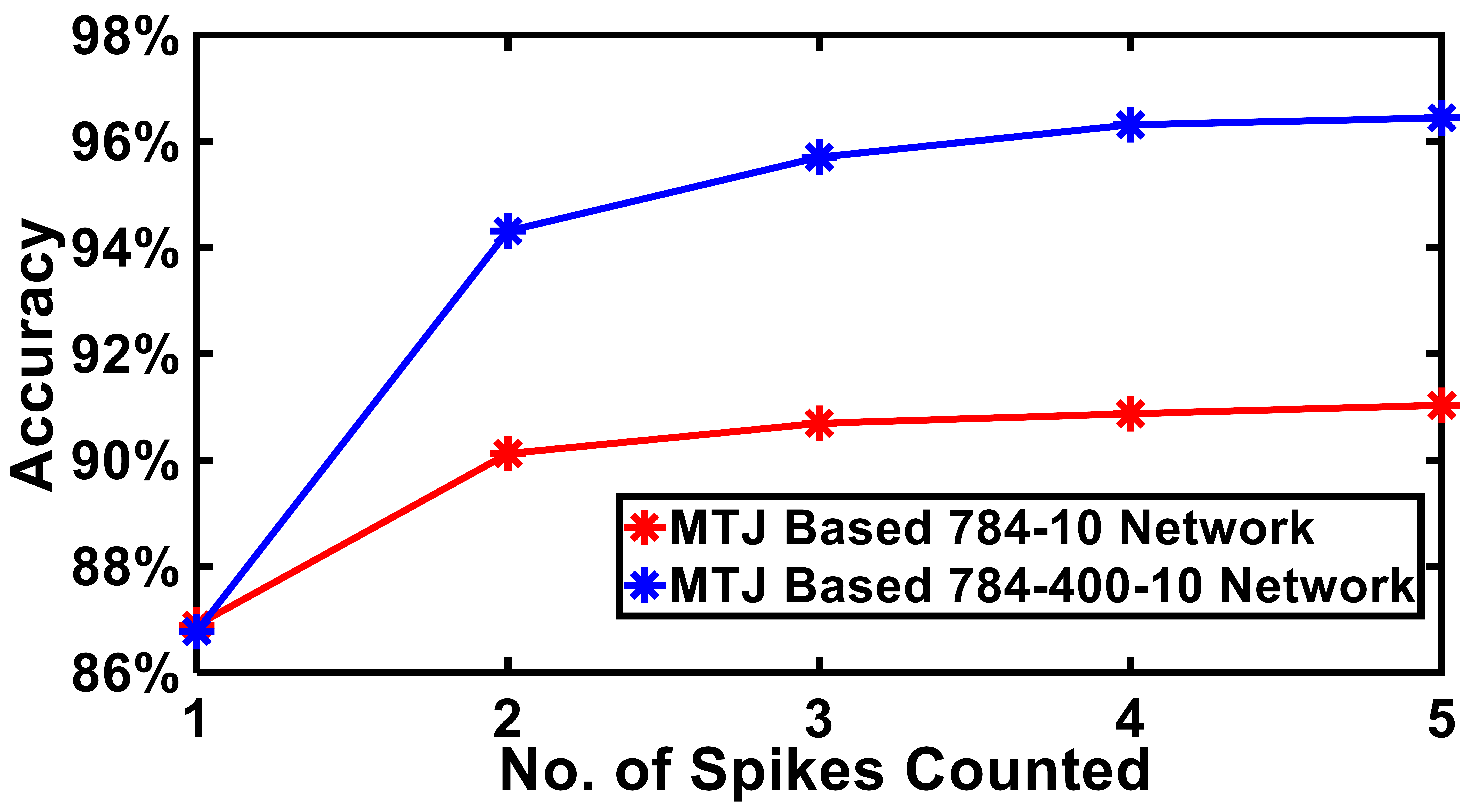}} 
  \caption{\textbf{Algorithm Formulations:} (a) Supervised algorithm for stochastic SNNs with temporal information encoding where neuron input, $V_2$, controls the time to fire. (b) Variation of the average device lifetimes as a function of the neuron input, $V_2$, which is equivalent to the weighted summation of synaptic inputs $\sum w_{i} I_{i}$. Device lifetime, $\tau_{AP}$, remains roughly constant over the input voltage range while the exponential variation of $\tau_P$ with $V_2$ is considered to be the activation function of the neuron ($g(.)$ in Eq. (4)). (c) Accuracy of MTJ based hardware simulations for two neural network architectures ($784 \times 10$ and $784 \times 400 \times 10$ neurons) are depicted. The $784 \times 400 \times 10$  ($784 \times 10$) network has a baseline accuracy of $97.41\%$ ($90.88\%$). Simulated accuracy of the hardware MTJ network approaches the baseline software accuracy with time-to-2nd/3rd spike of the winning neuron.} 
  \vspace{-2mm}
\end{figure*}

Given such a continuously switching device is available where the precise temporal dynamics can be controlled, the high level question to be addressed next is: \textit{Can we map the core device characteristics to compute primitives required in a functional stochastic SNN operation with temporal information encoding?} Let us consider a particular network where all the neurons are driven by the same voltage corresponding to input $V_1$ such that the average device lifetime in the AP state equals the duration of a ``timestep" in the system. Note that the duration of ``timestep" will be determined by circuit and architecture level constraints for simulating the SNN. If we interpret the device AP state as the ``spike" of the neuron, then the average time to fire for that neuron will be given by $\tau_P$, which can be controlled by the external neuron input $V_2$. For an SNN inferring data based on temporal encoding, this time to fire will dictate the winning neuron. The neuron which fires earliest will be interpreted as the winning class and is based on time-to-first-spike encoding. Note that the SNN can be turned off after the first spike, thereby resulting in significant sparsity and latency benefits. Such a fine-grained control of time to fire is not possible in case of stochastic magnetic devices driven by only a single external input signal since both the device lifetimes will be modulated together. It is also worth mentioning here that while our proposal is based on the ME-MTJ device, the formulation can be easily extended to experimentally demonstrated stochastic devices operating under the influence of external spin current and magnetic field \cite{zink2018telegraphic,zink2019independent}. In order to train the network, let us assume that we set the winning class neuron to fire at timestep $t_1$ while the other neurons target a firing time $t_2$. In order to infer with sufficient confidence margin, $\Delta t =t_2 - t_1$ should be reasonably high. Note that $\Delta t,t_1$ and $t_2$ are hyperparameters for our algorithm and user specified. In this work, we used a value of $t_1 = 1 ns$ and $t_2 = 300 ns$. 
\vspace{-2mm}

\section{Algorithm Formulation}
Fully connected neural network architectures with stochastic temporal encoding were trained on the MNIST dataset \cite{lecun1998gradient} based on algorithmic formulations described next. Since the real-time device lifetimes follow an exponential distribution in the low current regime \cite{Vincent2015}, we utilize Kullback-Leibler (KL) divergence to model the loss function. Assuming the target average device lifetime in the P state to be $\lambda$ and the expected device lifetime due to the external input to be $z$, the KL divergence between the expected and target spike probability distributions is given by,
\begin{equation}
L = \sum_{a\in{A}} \frac{1}{\lambda}e^{-\frac{a}{\lambda}}\log(\frac{z}{\lambda}e^{a(\frac{1}{z}-\frac{1}{\lambda})})
\label{Loss Function}
\end{equation}
where, $A$ is the probability space. From a network perspective, each neuron receives the weighted summation of synaptic inputs ($\sum_{i} w_iI_i$) as the input voltage $V_2$ (see Fig. 3(a)). Note that the output current in the spike detection circuit (see Fig. 2(b)) can be used to charge a capacitor till the input neuron device spikes, thereby converting the timing information to an analog voltage input for the next layer. Assuming the intrinsic device function mapping from the synaptic dot product to the average P state device lifetime to be $g(.)$ (which can be formulated by the exponential variation shown in Fig. 3(b)), 
\begin{equation}
z = g(\sum_{i} w_iI_i) = g(V_2)
\label{g function}
\end{equation}
It is worth mentioning here that the output $z$ represents the average value of P-state device lifetime under the influence of $V_2$, although the real-time characteristics follow an exponential distribution \cite{Vincent2015}. The operating voltage range of the device is also chosen properly (Fig. 3(b)) such that the change in $\tau_{P}$ is much larger than $\tau_{AP}$ (assumed constant equal to spike duration in the algorithm formulation) within this working range.

Using gradient descent, the weights of the network can be learnt through the following relations,
\begin{equation}
w = w-\alpha (\frac{\partial L}{\partial w}); \frac{\partial L}{\partial w} = \frac{\partial L}{\partial z}\frac{\partial z}{\partial w}
\label{gradient descent}
\end{equation}
where, $\alpha$ is the learning rate. The term $\frac{\partial z}{\partial w}$ can be obtained using Eq. (4), while the term $\frac{\partial L}{\partial z}$ can be derived from Eq. (3) by algebraic manipulations as, 
\begin{equation}
\frac{\partial L}{\partial z} = \sum_{a} \frac{1}{z\lambda}e^{-\frac{a}{\lambda}} -\sum_{a}\frac{a}{T^2\lambda}e^{-\frac{a}{\lambda}}
\label{plpz}
\end{equation}

The activation function of the neurons, given by the relationship between the P state lifetime, $\tau_{P}$, and the applied voltage, $V_{2}$, is obtained from stochastic-LLG simulations of the superparamagnetic MTJ device with a $2K_BT_K$ barrier height. A hybrid device-algorithm co-simulation framework calibrated to experimental measurements was used to evaluate the performance of the network. The $784\times 10$ network therefore consisted of LLG simulations of 10 MTJ devices while the deeper $784 \times 400 \times 10$ network consisted of 400 MTJs in the hidden layer and 10 devices in the output layer.

We observed a test accuracy of $90.88\%$ for a network architecture of $784 \times 10$ neurons. However, since the real-time device operation is stochastic with exponential lifetime characteristics, there might be image instances which are inferred incorrectly if the decision is solely based on the first spike. In that case, the robustness of the decision and the classification accuracy improves significantly if the inference process is based on the sum of multiple inter-spike intervals. As demonstrated in Fig. 3(c), the accuracy of the hardware network approaches the ideal baseline software accuracy with only a $2/3$-spike confidence for the winning neuron, thereby resulting in a highly sparse firing behavior of the neurons due to temporal information encoding. 

Similar observations were achieved when the network was scaled to a $3$-layer architecture with $784 \times 400 \times 10$ neurons. The network had a test accuracy of $97.41\%$, at par with iso-architecture standard deterministic networks (a conventional non-spiking network with rectified linear neuron units with 400 hidden layer neurons was observed to have a test accuracy of $97.03\%$ after 20 epochs of training). Interestingly, even for this deeper network, the testing accuracy achieved near-maximum values with only $2-3$ spikes being considered for both the hidden and output layers. This is a significant improvement over rate encoding methods and substantiates the advantages of spiking sparsity enabled by temporal encoding. In rate encoding, each layer triggers the next layer by the average firing rate and therefore the spiking activity increases exponentially with layer depth (for instance, the maximum firing activity per neuron can range between $5-10$ in deep rate encoded SNN architectures like VGG and ResNet \cite{sengupta2019going}). In contrast for temporal encoding, since information transmission from one layer to another does not depend on average firing rate but rather on the time of firing, there is no dependency of spiking activity on network scaling. While the stochasticity causes the number of spikes for inference to slightly increase above 1 to maintain minimal accuracy drop, it enables the usage of binary state-compressed scaled neuron devices to encode multi-bit information, instead of complex device structures exhibiting spin textures like domain walls, skyrmions, among others \cite{sengupta2017encoding}. In order to perform a benchmarking analysis, we compared the sparsity levels in our network against an iso-accuracy rate-encoded stochastic MTJ network (implemented according to the proposal outlined in Ref. \cite{sengupta2016probabilistic}). We observed $1.6\times$ reduction in spiking sparsity for the hidden layer and $3.77\times$ reduction in spiking sparsity for the output layer in the $784\times 400 \times 10$ neuron network. Scaling to deeper architectures is expected to improve the sparsity and latency benefits of such architectures along with providing accuracies at par with other implementations \cite{severa2019training,shrestha2018slayer}.

\section{Discussion and Outlook}
The article presents a unique perspective of designing efficient stochastic neuromorphic systems with temporal information encoding driven by an interdisciplinary perspective from devices to brain-inspired algorithm development. The work provides algorithmic formulations to leverage the stochastic temporal device characteristics of superparamagnetic devices and provides proof-of-concept demonstrations through extensive simulations. Such an end-to-end co-design effort to leverage unique properties of neuromorphic computing is an ideal fit for application drivers characterized by temporal information (for instance, sparse data collected by event-driven sensors \cite{singh2021gesture,mahapatra2020power}, among others).


\end{document}